\documentclass[onecolumn]{aa}
\usepackage{natbib}
\usepackage{txfonts}
\usepackage{psfig} 
\usepackage{graphicx}
\begin{document}
  \title{The  Dyer-Roeder   relation  in  a   universe  with  particle production} 
 \author{M.  de  Campos   \inst{1}  \and  J.  A.  Souza\inst{2}. }  
\institute{Physics Department, UFRR,Campus
   do    Paricar\~ana,   Boa    Vista,   RR.    Brasil.    \\   \email
   {campos@dfis.ufrr.br} \and  UFF Physics Institute,  Av. Litor\^anea
   s/n, Niter\'oi, RJ. Brasil.\\ \email{joseuff@bol.com.br}}
\abstract{  We  have  obtained   analytical  exact  solutions  of  the
Dyer-Roeder  equation  in  a  cosmological  model  where  creation  of
particles  occurs  at the  expense  of  the  gravitational field.   We
discuss the influences of inhomogeneities  in the path of a light beam
on the  apparent diameter of  astrophysical objects and  consider both
redshift  independent  and  redshift  dependent distributions  of  the
inhomogeneities.    
}
\maketitle
\section{Introduction}
Cosmology  has  been  ,  for   a  long  time,  a  fertile  ground  for
speculation.  The choice between competing theories was very difficult
due to the small amount of reliable experimental data.

Things have  changed, however, in  the last decades.  The  quantity of
experimental results  relating to the  age of universe,  its expansion
and its matter distribution,  as well as gravitational lens occurrence
statistics and related  subjects, has grown to such  a extent that the
room  left  today for  speculative  reasoning  in  Cosmology has  been
considerable reduced.

Among the most  interesting recent results, are the  supernova IA type
data, obtained  at the  end of  the 1990s, which  gave support  to the
hypothesis   that   our   universe   has  an   accelerated   expansion
\citep{Riess,Perlmuter}.

These observations lead to a  revival of the cosmological constant, as
well as  to new proposals for  candidates able to  generate a negative
pressure \citep{Caldwell,Peebles}, for example, quintessence.

According to some of these hypothesis, the universe would have, beyond
its usual  baryonic matter  content and dark  matter, also  a negative
pressure-generating content, a kind of dark energy that represents the
vacuum contribution \citep{Bharat,Carrol,Sahni}.

One  of  the  attractive   features  of  the  hypothesis  of  particle
production is that it  therefore relates the large-scale properties of
the universe to atomic  phenomena \citep{MacCrea}.  On the other hand,
the  introduction of  this new  component for  the  cosmological fluid
gives not  only an explanation for the  cosmological acceleration, but
also eliminates the age problem of the universe, which in the standard
model is  smaller than the  one obtained for  the age of  the globular
clusters.  The  estimate for the age  of the universe  depend upon the
value of the Hubble constant.  If the value of $H_0$ is near the upper
limit   obtained   by  Freedmann   et   al.    ($H_0  =   80km/s/Mpc$)
\citep{Freedman}  and  considering  the  usual standard  model  ($H  =
\frac{2}{3t}$), we get some "relief" for the age problem, although not
a definitive solution \citep{Ozer}.

The model  with particle production  (OSC) provides also  a reasonable
fit with  respect to kinetic tests, like  luminosity distance, angular
diameter and  the number counts  of galaxies versus  redshift relation
\citep{Alcaniz,Lima1}  and  in  the  radiation-dominated  era  (photon
creation) the  model can be  compatible with present day  isotropy and
the   spectral  distribution   of  Cosmic   Microwave  Background
Radiation \citep{Lima2}.

The inclusion of $\Lambda $ solves the age of the universe puzzle, but
at  the expense  of creating  a  new one,  the so-called  cosmological
constant  problem \citep{Weinberg}.  The  conciliation between  a very
large value for this constant,  predicted by quantum field theory, and
a small one  or zero, can be obtained if  we consider the cosmological
term time-dependent or quintessence  models. In spite of, these models
cannot  explain why  the dark energy  density is comparable  with the
matter one.

As  an  alternative  model  for  the  universe,  we  can  introduce  a
cosmological particle  production term, resulting in a  scenario that can
mimic the effects generated by the inclusion of $\Lambda$.
The physics involved is, nevertheless, quite different.

In this work we are going  to study the exact solutions of the Dyer-Roeder
equation,  considering  a  homogeneous  and isotropic  universe  where
particle  production occurs  at  the expense  of gravitational  field
energy \citep{Prigogine,Lima,Waga,Gariel,Zindhal,Lima1,Alcaniz}.

In  section 2  we  outline  the cosmological  model  and the
relation between the source of  particle production and the cosmological
term; in  section 3 we  examine the observational constraints  for the
model with  respect to the age  of the universe ,  the acceleration of
its   expansion,  and  make   some  considerations   about  primordial
nucleosynthesis  and structure formation,  obtaining a  narrower range
for the  values of  the creation particle  parameter; in section  4 we
obtain  the   Dyer-Roeder  equation  for  a   universe  with  particle
production  and   find  exact  solutions   considering  both a redshift
independent and redshift dependent distribution of inhomogeneities.
\section{The model}
\subsection{Field equations}
In this section  we give a summary  of the cosmology with
particle production \citep{Prigogine} ,
show how  this model mimics  the universe with  a $\Lambda $  term and
find  constraints in  the  creation  parameter using  the  age of  the
universe and the deceleration parameter.

  The universe will be considered homogeneous and isotropic, described
by the FRW line element
\begin{equation}
              ds^{2}=dt^{2}-R^{2}(t)[dr^{2}+r^{2}d               \theta
              ^{2}+r^{2}sin^{2}(\theta)d \phi ^{2}] \, .
                      \end{equation} 

The  model  with  particle  production  is  obtained  from  the  field
equations
\begin{eqnarray}
8\pi  G\rho &=&  3\frac{\dot{R}^2}{R^2}+3\frac{\kappa }{R^2}  \,  , \\
8\pi                       (P_{th}+\tilde{P})                      &=&
-2\frac{\ddot{R}}{R}-\frac{\dot{R}^2}{R^2}-\frac{\kappa }{R^2 } \, ,
\end{eqnarray}
coupled to the balance equation for the particle number density
\begin{equation}
\dot{n}+3\frac{\dot{R}}{R}n = \Psi\, .
\end{equation}
Here  the dot  means time  derivative, $\rho$  is the  energy density,
$P_{th}$  is  the  thermodynamical  pressure,  and n  and  $\Psi$  are
respectively  the  particle  number  density  and the  source  of  the
particle  production.   $P_c$  is  the  creation  pressure,  given  by
\citep{Calvao}
\begin{equation}
\tilde{P} = -\frac{\rho + P_{th}}{3nH} \Psi \, ,
\end{equation}
when the particle production is considered as adiabatic process.

Combining equations (2), (3) and  (4) with the state equation $P_{th}=
(\nu-1) \rho$, it follows that
\begin{equation}
R\ddot{R}+(\frac{3\nu  -2}{2}-\frac{\nu  \psi  }{2nH})( \dot{R}  ^2  +
\kappa) = 0 \, ,
\end{equation}
where $\nu $ is a constant and $\kappa $ is the curvature.

For  a complete description  of the  cosmological environment  we must
have  an explicit expression  for the  source of  particle production.
Following Lima et al. \citep{Lima}, a physically reasonable expression
for  the particle  creation  rate is  $\Psi =  3  n \beta  H$.  {  The
creation parameter $\beta $ is generally a function of the cosmic era,
or equivalently of the parameter $\nu $.  We can assume two
different parameter  creations, $\beta _{\nu }$ and  $\beta _{m}$, for
radiation  and matter  eras,  respectively \citep{Alcaniz}.   However,
since we study a distance treatment in the actual
matter dominated phase ($\nu =0 $), we supposed $\beta _{m} = \beta $,
with  $\beta$ constant.}   Beyond  this, $\beta  $  must be  positive,
otherwise  the  second  law  of  thermodynamics  (${S^{\alpha}}  _  {;
\alpha}> 0 $) is violated.

Substituting the  particle production rate  above in equation  (6) and
integrating , we obtain
\begin{equation}
  R = R_{0}(\frac{t}{t_{0}})^{\frac{2}{3\nu (1-\beta)}} \, ,
 \end{equation}  
where the subscript $0$ refers to the present time.

The OSC can mimic a universe  with a $\Lambda $ term. To clarify this
relation we  take into account  the entropy production in  both cases.
For OSC we have
\begin{eqnarray}
S^{\alpha  }_{;  \alpha} =  n\dot{\sigma}  +\sigma  \psi =  -\frac{\mu
\psi}{T} \, , \nonumber
\end{eqnarray}
while for a universe with a $\Lambda $ term
\begin{eqnarray}
S^{\alpha  }_{;   \alpha}  =  -\frac{\dot{\Lambda  }}{8\pi   G  T}  \,
.\nonumber
\end{eqnarray}
$T $ is  the temperature, $\mu = \frac{\rho + P}{n}  -T\sigma $ is the
chemical potential and $\sigma $ is the entropy per particle.

The two  scenarios can  explain the entropy  content of  the universe,
which is poorly explained using only Einstein's field equations,
which are invariant under time reversal \citep{Prigogine}.

Considering  the  process to be  adiabatic   ($\dot{\sigma  }  =  0  $)  and
equalizing  the specific entropy  due to the $\Lambda $  term and  due to
particle production we obtain
\begin{eqnarray}
\frac{\dot   {\Lambda}}{8   \pi   G}   \propto   \frac{\rho   \psi}{n}
\,. \nonumber
\end{eqnarray}

For $\psi  = 3n\beta  H$ this  implies
$\Lambda  \propto  t^{-2}$,  for  a  pressurelles  universe.   The  two
scenarios  have identical time  dependence for  the scale  factor, and
consequently for the Hubble function, too.
\subsection{Observational constraints}
\subsubsection{Age of the universe}
Using the scale factor (7) the Hubble function is given by
\begin{equation}
H = \frac{2}{3}[(1-\beta)t]^{-1} .
\end{equation}
The   results   from   the   Hubble  Space   Telescope   Key   Project
\citep{Freedman2} indicate,
\begin{equation}
H_0 = 72 \pm 8 \frac{Km}{Mpc \cdot s} \, ,
\end{equation}
for the Hubble constant.

The High Z  Supernovae estimates for the  age of the 
universe is 14.2 $\pm  $ 1.7 Gyrs \citep{Riess} and 14.9  $\pm $ 1.4 Gyrs
\citep{Perlmuter}.  Carreta et  al.(2000) using results from
Hipparcos, RRLyrae and Chepheids to re-calibrate the globular cluster
distance scale,  found 12.9 $\pm $ 2.9  Gyrs.  Butcher (1987)
used the abundance of the $Th^{32}  $ to estimate the age
of  the stars  and  consequently a  lower  limit for  the  age of  the
universe.  The  interest in  $Th$ abundance is  related to  its half
life, about 14.05  Gyr, an interesting time from  a cosmological point
of view.   Using the relation between  the abundance of $Th  $ and $Eu$,
Westin et al. (2000) estimate  for the universe an age of 15.0
Gyr,  while Johnson  and  Bolte (2000)  found 11.4  Gyr.
Recently Krauss  and Chaboyer (2000) estimates a lower
limit on the age of the universe, 11 Gyr with $95 \% $ confidence.

These values
differs substantially from each  other, depending on the technique
used; we  update the interval for  the value of  the $\beta$ parameter
found by Lima et al. \citep{Lima}.

Using (8),  the upper limit  for $H_0 $  from Freedman et al.  and the
estimated age for  the oldest objects in our galaxy  ($16.3 Gyr $), we
obtain  the value  $  0.49 $  for  the upper  limit  for the  creation
parameter $\beta  $.  To obtain the  lower limit for $\beta  $, we use
the same  expression (8), but with  the lower limit for  $H_0 $, again
from Freedman,  and the smallest age  for the local  oldest stars from
Carreta et al. (2000), giving $0.02$.

So, the validity interval for the creation parameter is
\begin{eqnarray}
0.02 < \beta < 0.49 \,. \nonumber
\end{eqnarray}

\subsubsection{Acceleration of the universe}
According to  data from type Ia supernovae  observations, the universe
is accelerating.  Generally, this acceleration is thought to be due to
 repulsive gravity,  which can be introduced via a negative
pressure,  a cosmological  term  or a  quintessence  component in  the
matter content of the universe.

An  alternative view  would be  to consider  particle  creation, which
naturally redefine  the energy-momentum  tensor and could  account for
this  increasing expansion  velocity  \citep{Freaza,Alcaniz,Lima,Balakin}.

Taking into account (6), we  can write the deceleration parameter in
terms of the particle creation source, namely
\begin{equation}
Q =\frac{1}{2}-\frac{4\pi m G \Psi}{H^3}\, ,
\end{equation}
where $m $ is the rest mass of the produced particles, and we consider
a null  value for the  curvature.  We obtain an  accelerated universe,
expanding with the scale factor (7), if $\beta > \frac{1}{3}$.  Taking
into account the age of the oldest objects in our galaxy, the estimate
for  the  Hubble constant  and  the indications  that  we  live in  an
accelerating universe, we obtain that
\begin{eqnarray}
 0.33 \leq \beta \leq 0.49 \, . \nonumber
\end{eqnarray}
\subsection{Primordial nucleosynthesis}
In spite  of the cosmological  consequences, the current matter  creation rate
, $\Psi_0  \approx 10^{-16}$ nucleons $cm ^{-3}  yr^{-1}$ , is
nearly  the   same  rate   predicted  by  the   steady-state  Universe
\citep{Hoyle} taking into account  $\kappa = 0$.  This matter creation
rate     is     presently      far     below     detectable     limits
\citep{Alcaniz}.  Consequently,  this  indicates, probably,  that  the
early picture of  the OSC differs slightly from  the standard model with
respect  to the  primordial nucleosynthesis.   However  a quantitative
analysis  with a multifluid description  is necessary to more
accurately determine the primordial nucleosynthesis in OSC.

Considering  primordial nucleosynthesis  in its  standard form  with 3
neutrinos  flavours, the  main parameter  of  the theory  is the  baryon
number density quoted  as its ratio to the  number density of photons,
namely $\eta  = \frac{n_b}{n_{\gamma}} $.  Note that,  in a given
era of evolution of the universe, the model with cosmological particle
production affects the number density  of photons in the same way that
the number  density of the baryons.   So, in this model  $\eta$ does not
suffer sensitive variations in relation to the standard model.
\subsection{Growth of structures}
In the OSC  model the particles  are created from the  gravitational field
energy.  In this case, a  negative extra pressure appears in the energy
momentum tensor.  This extra pressure is  responsible for cosmic
acceleration and has the same influence as the negative pressure that
appears in  models with a  cosmological term.  So, in  the same way
that the effects of a cosmological term on linear perturbation theory
over short distances can be studied using the expression
\begin{eqnarray}
\ddot{\delta} + 2H\dot{\delta}-4\pi G \rho \delta =0 \, , \nonumber
\end{eqnarray}
we consider  that this equation is  also applicable to  the OSC model,
where  $\delta $  is  the density  contrast.   Substituting the  scale
factor  (7) in  the equation  for  the density  contrast evolution  we
obtain the following modes
\begin{eqnarray}
\delta _-  &\propto & t^{\frac{3\beta +1+\sqrt{9 \beta  ^2 +25 +6\beta
}}{6(\beta -1)}}  \nonumber \\  \delta _+ &\propto  & t^{\frac{-3\beta
-1+\sqrt{9 \beta ^2 +25 +6\beta }}{6(1- \beta )}}\, . \nonumber
\end{eqnarray}
For $\beta  = 0 $ we obtain  the usual growing and  decaying modes for
the  dust  Friedmann  model,  $\delta _  +  \propto  t^{\frac{2}{3}}$,
$\delta _ -  \propto t^{-1}$, respectively.  Note that  the $\delta _
+$  for the  OSC  model grows  faster  than the  growing mode  without
particle production ($\beta  =0$), including for accelerated expansion
of the  universe.  The few indications  in this section  are, from our
point of  view, sufficient to motivate  a more detailed  study of the
scalar density perturbations to furnish a more precise picture of the growth
 of structures in the OSC model.
\section{The Dyer-Roeder equation}
If there is matter distribution along  a cone defined by the bundle of
light  rays connecting the  source to  observer, the  angular diameter
distance of that  source from the observer is  smaller than that which
would occur if the source was seen through an empty cone.  This effect
is known as  Ricci focusing. Our intention in this  section is to find
and solve the mathematical  expression of this effect, the Dyer-Roeder
equation, when the particle number in the universe is not conserved.

For a conformally flat metric the Ricci focalization equation is
\begin{equation}
\ddot{\sqrt{A}}+\frac{1}{2}R_{\alpha
\gamma}k^{\alpha}k^{\gamma}\sqrt{A}=0 \, ,
\end{equation}
where the dot  means derivative with respect to  the affine parameter,
$A  $ is  the beam  cross  sectional area  and $k^{\alpha}=  \frac{dx^
\alpha}{dv}  $  is  the   vector  tangent  to  the  photon  trajectory
\citep{Demianski}.  Considering an energy momentum tensor of the form
\begin{equation}
T_{\mu \nu }=(\rho +P )u_{\mu}u_{\nu} +Pg_{\mu \nu} \, ,
\end{equation}
where   the  total   pressure  P   includes  contributions   from  the
 thermodynamical  and   creation  pressures,  we   obtain,  after  
 straightforward algebra \citep{Amendola}
\begin{equation}
\ddot{\sqrt{A}}+4\pi\frac{\rho + P}{H_0 ^2}(1+z)^2 \sqrt{A}=0 \, .
\end{equation}
Taking into account  equations (5) and (7), we  obtain, after a simple
but  a lengthy  manipulation \citep{Amendola},  the Dyer-Roeder
equation in the cosmological background with particle production
\begin{equation}
(1+z)^5                       \frac{d^2                       r}{dz^2}
+\frac{7}{2}(1+z)^4\frac{dr}{dz}+\frac{3}{2}(1-\beta)[1+z]^{3(1-\beta)}
\alpha r =0 \, ,
\end{equation}
where we  have defined the  dimensionless distance $r  = D H_0$,  $D $
as the  angular diameter  distance.  The  parameter  $\alpha $  is
related to the Ricci focusing  and conceptually is the  fraction of
homogeneously distributed matter inside the light beam \citep{Roeder}.

Let us consider two interesting particular cases.

\subsection{Constant $\alpha$}
We  recall that, for  $\alpha =  0$, our  model represents  a universe
where all  matter is  clustered, while,  for $\alpha =  1$, we  have a
universe with its matter content homogeneously distributed.

Equation (14) can be analytically solved for any value of $\alpha $ in
the interval from zero to one.  We obtain
\begin{equation}
r(z)  =   (1+z)^{-5/4}\{C_1  J_{\frac{\pm  5}{6   \beta}}[\Delta  ]  +
C_2Y_{\frac{\pm 5}{6 \beta}}[\Delta ]\} \, ,
\end{equation}
where
\begin{equation}
\Delta    =   (\frac{2\alpha    (1-\beta)}{3\beta   ^2})^{\frac{1}{2}}
(1+z)^{\frac{-3\beta }{2}} \, .
\end{equation}
$J$  and  $Y$  are  the   first  and  second   Bessel  functions,
respectively.

The profile for $r(z)$ when  the universe is non accelerated ($\beta <
\frac{1}{3}$) is shown in Fig. 1.
\begin{figure}[!ht]
\centerline{\includegraphics[width=8cm]{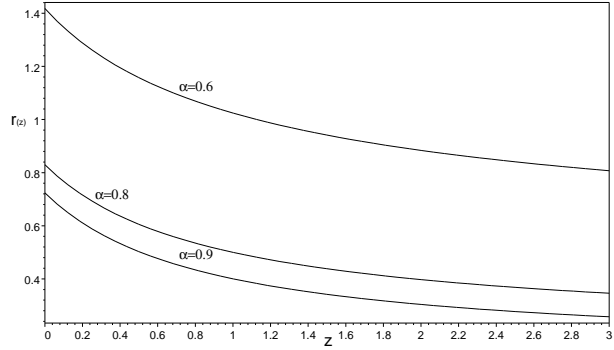}}
\caption{Evolution  of Dyer-Roeder  distance  for $\beta  = 0.1$,  and
three different values of $\alpha$.}
\label{fig:Fig. 1}
\end{figure}

For an accelerated universe  ($\beta > \frac{1}{3}$)
the profile is the one seen in Fig. 2.

\begin{figure}[!ht]
\centerline{\includegraphics[width=8cm]{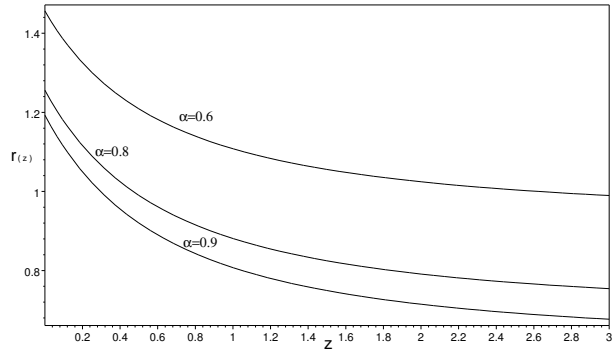}}
\caption{Evolution of Dyer-Roeder distance  for $\beta = 0.4$, and the
same three different values of $\alpha$.}
\label{fig:Fig. 2}
\end{figure}
To obtain  the limiting case when  there is no  particle production we
must put $\beta =0 $ in the original differential equation (14).
\subsection{Variable $\alpha $}
In this section  we solve the Dyer-Roeder equation  for the case where
$\alpha $ is no longer a  constant, but instead a particular function of
the redshift.  We  adopt for $\alpha (z) $  an expression identical to
the one used by Linder (1998), namely
\begin{equation}
\alpha (z) = p (1+z)^{q } \, ,
\end{equation}
where $p  $ and $q$ are  constants.  It is reasonable  to expect the
clumping to  decrease with increasing  redshift, implying $q  > 0$.
Some care is needing in choosing the parameters $p$ and $q$ is 
due to the range of $\alpha$,  that must stay  in the interval  $0 \leq
\alpha \leq 1$.  Naturally, this  will depend on the range of redshift
studied.

Integration of equation (14) now results in
\begin{equation}
r(z)={\it \_C1}\,{\it J}_{A}(B)+{\it \_C2}\,{\it Y}_{A}(B) \, ,
\end{equation}
where
\begin{eqnarray}
A =\frac{5}{2\,\left (5-3\,\beta+q\right )} \nonumber
\end{eqnarray}
and
\begin{eqnarray}
B={\frac       {\sqrt        {6(1-\beta       )p}       (1+z       )^{
5/2-3/2\,\beta+1/2\,q}}{5-3\,\beta+q}} \, .  \nonumber
\end{eqnarray}
The profile of  $r(z)$ in this case, is given by  Fig. 3, for specific
values of  $q$ and $p$, when  the universe is  not accelerated ($\beta
<\frac{1}{3}$).
\begin{figure}[!ht]
\centerline{\includegraphics[width=8cm]{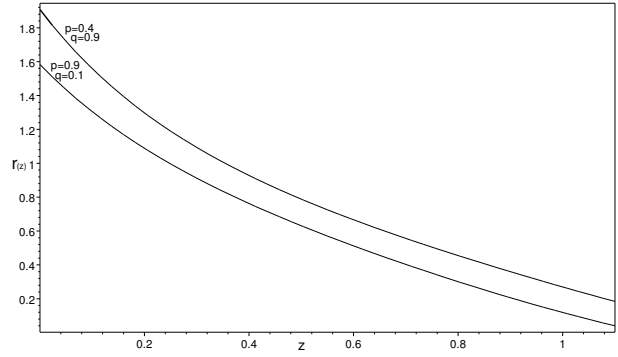}}
\caption{Evolution of the Dyer-Roeder distance for $\beta = 0.25$, and
  two different values of $q$ and $p$.}
\label{fig:Fig 3}
\end{figure}

For  an accelerated universe  ($\beta >  \frac{1}{3}$) the  profile is
illustrated in Fig. 4.
\begin{figure}[!ht]
\centerline{\includegraphics[width=8cm]{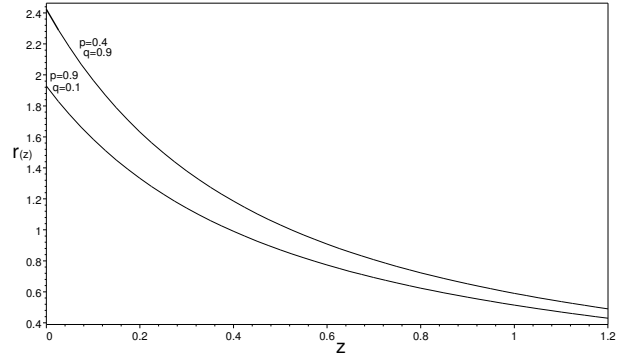}}
\caption{Evolution of Dyer-Roeder distance for $\beta = 0.8$, and same
 two different values of $q$ and $p$.}
\label{fig:Fig. 4}
\end{figure}
\newpage
\section{Conclusions and final remarks}
The current matter creation rate, $\Psi_0 \approx 10^{-16}$ nucleons
$cm  ^{-3}  yr^{-1}$  , is  nearly  the  same  rate predicted  by  the
steady-state  universe  \citep{Hoyle},  regardless of  the value  of  the
curvature  parameter.   In  spite  of  the  matter  creation  rate
presently being far below detectable  limits \citep{Alcaniz},  the physical
consequences  for  the  evolution  of  the  universe  are  potentially
measurable.

We have updated
 (and narrowed) the range of allowed values for the creation parameter
 $\beta$.

For the Dyer-Roeder equation, we have  solved it in a
 background universe with  particle production, expanding considerably
 the  manifold of solutions.   As far  as we  know, this  equation has
 previously  been solved  only  for some  special  constant values  of
 $\alpha $ \citep{Amendola}.

In Fig. 1  we have plotted $r(z)$ for a  non accelerated universe, for
three  different  values  of   constant  $\alpha  $.   We  note  that
increasing the inhomogeneities in the  path of the light beam (smaller
$\alpha $) leads to greater values for the Dyer-Roeder distance at the
same redshift.  Remembering the  definition of $r(z)$, this implies
smaller values for the viewing angle as $\alpha $ diminishes.

We can  interpret these results 
 saying  that the objects will appear  at apparent distances
greater  than the  ones  we  would observe  in  a totally  homogeneous
universe ($\alpha = 1$).

In  Fig.  2 we  repeat  the  same graph,  but  now  in an  accelerated
universe.   We observe the  same qualitative  behavior, with little
enhancing  of the effect.   
The origin  of   this  effect  is   in  the  inhomogeneities,   while  the
acceleration, although having some influence, is not the main agent in
this case.

In Figs. (3) and (4) we show $r(z)$ in the case of variable $\alpha $,
given  by  (19), for  non  accelerated  ($\beta =0.25  $)  and
accelerated $(\beta = 0.8)$ universes, for different values of $q$ and
$p$.  In both cases, for  the redshift interval studied, the range of
the  smoothness function $\alpha(z)$ lies in the interval  $0 \leq
\alpha (z) \leq 1$.

The next  steps for future  work are to analyze the  lensing probability
occurrence for OSC.
%
%
\bibliographystyle{aa} \bibliography{ms3089}
\end{document}